\documentclass[sigconf]{acmart}
\usepackage{svg}
\usepackage{comment}
\usepackage{footmisc}

\AtBeginDocument{%
  \providecommand\BibTeX{{%
    \normalfont B\kern-0.5em{\scshape i\kern-0.25em b}\kern-0.8em\TeX}}}

\copyrightyear{2020} 
\acmYear{2020} 
\setcopyright{acmcopyright}\acmConference[KDD '20]{Proceedings of the 26th ACM SIGKDD Conference on Knowledge Discovery and Data Mining}{August 23--27, 2020}{Virtual Event, CA, USA}
\acmBooktitle{Proceedings of the 26th ACM SIGKDD Conference on Knowledge Discovery and Data Mining (KDD '20), August 23--27, 2020, Virtual Event, CA, USA}
\acmPrice{15.00}
\acmDOI{10.1145/3394486.3403380}
\acmISBN{978-1-4503-7998-4/20/08}

\newcommand{\IcM}{IMS}
\newcommand{\TransferAssistant}{DeepTriage}
\def\comment{}

\includecomment{realnumber}

\begin{document}
\fancyhead{}

\title{{\TransferAssistant}: Automated Transfer Assistance for Incidents in Cloud Services}

\author{Phuong Pham, Vivek Jain, Lukas Dauterman, Justin Ormont, Navendu Jain}
\affiliation{\institution{Microsoft}
  \city{Redmond}
  \state{Washington}
  \country{USA}
}
\email{phuong.pham, vivek.jain, lukas.dauterman, justin.ormont, navendu@microsoft.com}


\begin{abstract}
  As cloud services are growing and generating high revenues, the cost of downtime in these services is becoming significantly expensive. To reduce loss and service downtime, a critical primary step is to execute incident triage, the process of assigning a service incident to the correct responsible team, in a timely manner. An incorrect assignment risks additional incident reroutings and increases its time to mitigate by 10x. However, automated incident triage in large cloud services faces many challenges: (1) a highly imbalanced incident distribution from a large number of teams,
  (2) wide variety in formats of input data or data sources, 
  (3) scaling to meet production-grade requirements, and 
  (4) gaining engineers' trust in using machine learning recommendations. 
  To address these challenges, we introduce {\TransferAssistant}, an intelligent incident transfer service combining multiple machine learning techniques -- gradient boosted classifiers, clustering methods, and deep neural networks -- in an ensemble to recommend the responsible team to triage an incident. Experimental results on real incidents in Microsoft Azure show that our service achieves $82.9\%$ F1 score. For highly impacted incidents, {\TransferAssistant} achieves F1 score from $76.3\% - 91.3\%$. We have applied best practices and state-of-the-art frameworks to scale {\TransferAssistant} to handle incident routing for all cloud services. {\TransferAssistant} has been deployed in Azure since October 2017 and is used by thousands of teams daily.
\end{abstract}

\comment{


}

\maketitle

\section{Introduction}
Cloud services have become increasingly popular and are expected to gain $\$331.2$ billion in 2022, a $12.6\%$ compound annual growth rate since 2018 \cite{gartner2022}. With this impressive growth, any unplanned service interruption or outage risks customer dissatisfaction as well as a huge economic loss, e.g., $\$2.5$ billion every year for the Fortune 1,000 \cite{appdynamics}. Amazon is estimated to have a $\$100$ million loss due to its 2018 Prime Day outage \cite{prime}. 

Service interruptions can be detected by service monitors (creating live site incidents or LSIs) or reported by customers (creating customer reported incidents or CRIs). 
In a cloud setting, the incident life cycle is managed via an incident management system ({\IcM}). The fundamental step after incident creation is to route it to the responsible team, called incident triage, to minimize the service downtime. If the assignment is incorrect, it risks a high Time To Mitigate (TTM) as the incident gets rerouted from one team to another while customers are being impacted. Chen et al. \cite{chen2019empirical} found that $4.11\%-91.58\%$ of incidents were rerouted which increases the TTM to more than 10x 
in a large commercial cloud.

There are four key challenges making incident routing at cloud-scale fundamentally hard:
\begin{itemize}
    \item Large cloud platforms such as Amazon Web Services and Azure comprise tens of thousands of teams, typically organized by feature area or service, with complex inter-dependencies across them. Therefore, a customer issue such as 'virtual machine (VM) not responding' can be due to either the failed compute instance, network being disconnected, VM storage account being down, firewall mis-configuration, etc. 
    Further, this large number of teams poses extreme incident data imbalance as new teams will likely have a smaller scale and lesser incidents compared to older services \cite{matter2009assigning}.
    \item Running a cloud service generates a wide variety of monitoring data such as logs, stacktraces, performance counters 
    and service-level indicators. Further, during incident triage, engineers may run diagnostics, execute queries or  commands, and discuss the problem over IM or email which are logged in the incident. Therefore, routing the incident requires analyzing such diversity in types of input data and data sources, similar to challenges in bug triage~\cite{bhattacharya2012automated, chen2019continuous}. 
    Further, the content of an incident varies over time as engineers keep adding more information until it is mitigated \cite{chen2019continuous}. 
    \item Cloud services use agile development to continuously roll out service updates. Similarly, machine learning (ML) model deployments typically follow the same approach in quickly rolling out simple models first, getting user feedback and then building more sophisticated models to incorporate it. Therefore, we need to jointly optimize DeepTriage's accuracy along with the execution time and training cost to continuously deliver business value.
    \item Finally, engineers, responsible for mitigating incidents, are often used to applying deterministic rules to assign incidents. Given a ML recommendation, engineers have a common ask on how the model came up with the suggestion. Since it is fundamentally challenging to interpret most non-linear ML models \cite{choi2016retain, vondrick2013hoggles}, it is important to bridge this gap to gain engineers' trust to make ML-driven routing decisions. 
\end{itemize} 
While prior work focused on fast incident detection 
\cite{xu2009detecting, cohen2004correlating, cohen2005capturing, bodik2010fingerprinting}, there is little research on automating the incident assignment task \cite{chen2019continuous}. There has been
significant work in triage for software bugs \cite{matter2009assigning,lou2013software,bhattacharya2012automated}.
Some industry approaches provide a rule-based incident triage \cite{bigpandas} but rules can be hard to manage and do not automatically evolve with services.
Overall, the complex service dependencies, diversity of data and focus on fast mitigation pose new fundamental challenges for incident triage. 

In this paper, we introduce {\TransferAssistant}, an intelligent incident triage service running in Azure since 2017. {\TransferAssistant} combines state-of-the-art machine learning techniques  – gradient boosted classifiers, clustering methods, and deep neural networks – in an ensemble to handle different types of incident data. To make {\TransferAssistant} run in production, we have applied various best practices in the design and implementation to scale the service on-demand, ensure reliability, and handle the data distribution shift \cite{lee2017applying}. These techniques include data sampling and partitioning for class imbalance and reduce training times, model partitioning, performing training and inference steps in parallel, deploying models as cloud web services, decoupling components of the ensemble model for agile iterations, and systematic refresh and validation of the models. 

{\TransferAssistant} provides a REST endpoint which is leveraged by customers in various scenarios: (1) finding the right team when user is creating CRIs, (2) full-automated routing for LSIs, (3) integrated in {\IcM} user interface to provide triage recommendations to engineers, (4) automation workflows that combine rule-based triage with ML recommendations. 
Currently {\TransferAssistant} serves thousands of teams daily. Evaluation on hundreds of thousands of incidents show that our approach achieves F1 score equal to $82.9\%$ overall and $76.3\% - 91.3\%$ in high impacted incidents.
Model ablation analysis showed that each of the ML models we used provided a lift in the final ensemble for different incident types. To the best of our knowledge, we are the first one to present a deployed incident triage service for cloud-scale online services.

This paper makes three key contributions:
\begin{itemize}
    \item We introduce {\TransferAssistant}, the first deployed incident triage service  based on an ensemble of multiple machine learning techniques to recommend the responsible team for an incident. Experimental results on real data show that our model achieves good performance overall as well as in highly impacted situations.
    \item We share the best practices and technologies to enable {\TransferAssistant} meeting production requirements at cloud-scale.
    \item We present lessons learned and implications for deploying an incident triage system to drive down TTM in large online service systems.
\end{itemize}

This paper is organized as follow: Section \ref{sect:background} presents the background of an incident management system; Section \ref{sect:overview} provides details of   {\TransferAssistant}; Section \ref{sect:experiment} shows experimental results; Section \ref{sect:production} describes the deployment of {\TransferAssistant} in Azure; Section \ref{sect:discussion} discusses lessons learned and implications for implementing and deploying an incident triage service at cloud scale; 
Section \ref{sect:relatedWork} presents related work; 
and Section \ref{sect:conclusion} concludes this paper.

\section{Background} \label{sect:background}
\subsection{Incident Triage in {\IcM}}
A cloud service contains one or more services and each service has many teams. Each incident in an {\IcM} will be assigned to a specific team. An incident life cycle goes through four stages: incident creation, incident triage, incident mitigation and incident resolution. Incident creation could be either via a service monitoring alert 
or reported by customers. Incidents can have different severity levels e.g., 0-4, where severity 0 represents the highest priority incident having customer impact
whereas severity 4 incidents have low impact and typically 2-3 day SLAs and do not need to be dealt with immediately. {\IcM} maintains a roster of on-call engineers (OCEs), who are called or paged if a high priority incident (e.g., Sev 0-2) is received by the {\IcM}. In general, there is one primary OCE and one secondary OCE on the roster. In case the primary OCE does not answer or acknowledge the high priority incident, the {\IcM}
calls the secondary OCE. OCEs then acknowledge these incidents and begin their investigation.

\begin{figure}[h]
  \centering
  \includegraphics[width=\linewidth]{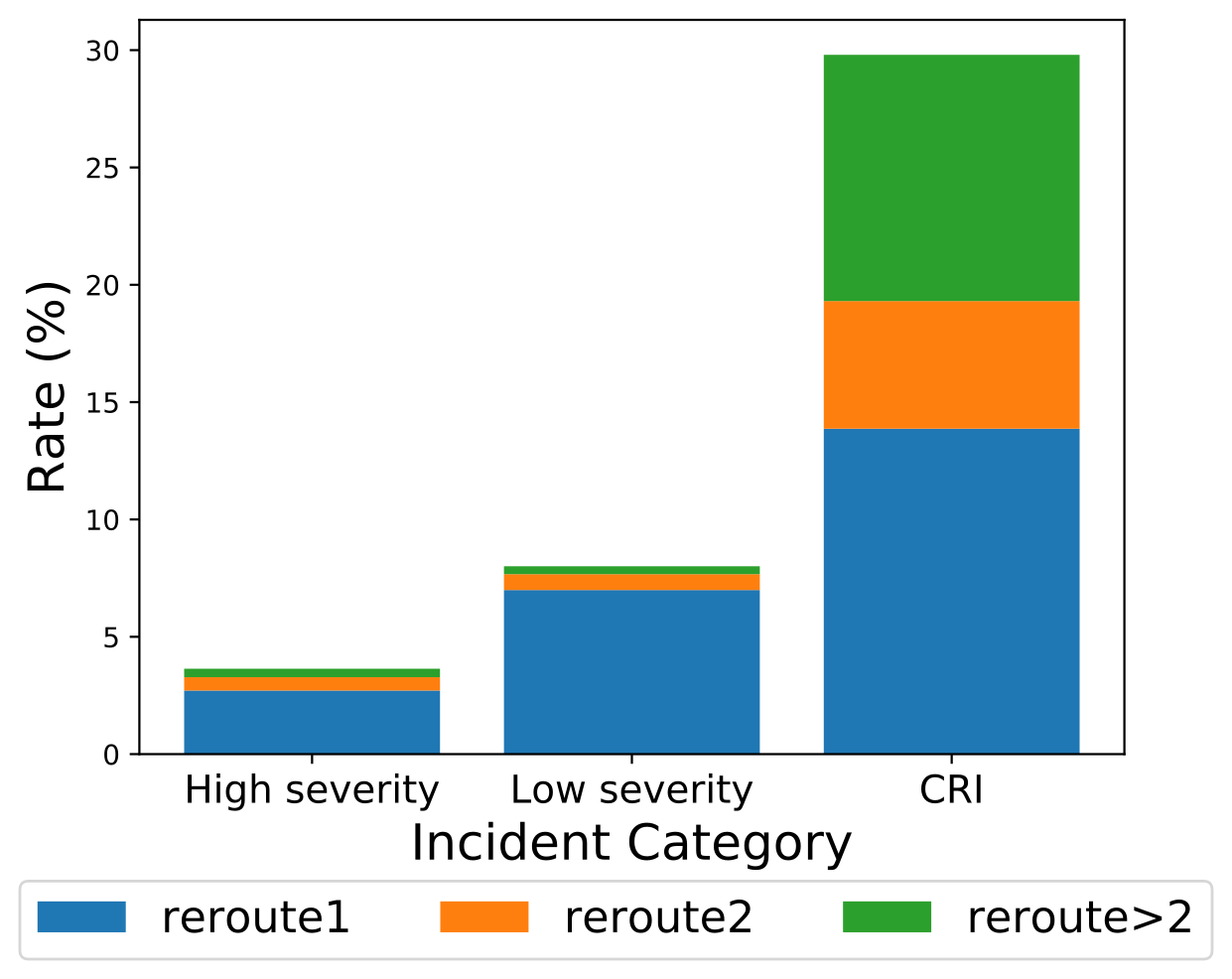}
  \caption{Incident rerouting rate in different scenarios}
  \Description{Rerouting Distribution of Incidents.}
  \label{fig:rerouteDist}
\end{figure}

Due to highly complex service dependencies, 
often incidents are not assigned to the right owners as there are a large number of possible teams (e.g., hundreds or thousands) to which an incident could be assigned. This causes incidents to be mis-routed and repeatedly reassigned across teams before an incident reaches the correct owning team.  Figure \ref{fig:rerouteDist} shows the rerouting distribution in three scenarios: high severity incidents (0-2), low severity incidents (3-4), and CRIs. We further categorized rerouted incidents into three buckets: only 1 reroute, having 2 reroutes, and more than 2 reroutes. While most rerouted incidents has 1 hop, CRIs pose a big challenge for {\IcM} because nearly one third of the CRIs got rerouted and many of them have more than 2 hops\footnote{Due to the policy of Microsoft, we cannot disclose the actual number of incident reports in this paper.\label{disclosure}}.

\subsection{Key Challenges}
There are four key challenges in building an automated incident routing system in cloud deployed {\IcM}.
\begin{itemize}
    \item A large number of possible teams, e.g. tens of thousands of teams, not only challenges machine learning models in terms of complexity but also exhibits extreme data imbalance. Figure \ref{fig:team_dist} shows the ratio of incident distribution of more than 10K teams in Azure over 6 months. While majority teams receive hundreds of thousands of incidents, there are some teams that have only one incident in the same time frame\footref{disclosure}.
\begin{figure}[h]
  \centering
    \includegraphics[width=\linewidth]{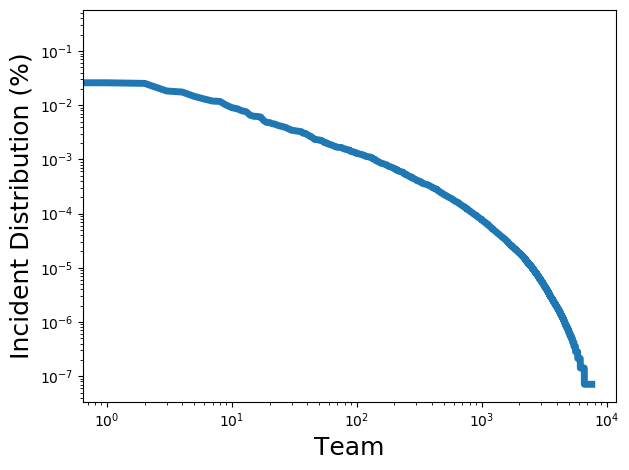}
  \caption{Number of incidents per team (log-scale).}
  \Description{Team Distribution}
  \label{fig:team_dist}
\end{figure}
\item A wide variety of data formats supported in modern {\IcM}s, e.g. HTML content, images, or logged IM, require the routing system to use different approaches to parsing this content and provides useful features information to machine learning models. Figure 3 shows the Venn diagram of incidents having at least a command line, or a query, or a stack trace content. We see that the incidents contain mixed data formats and an automated system cannot focus on a single format. Further, the content of an incident, changing over time as engineers keep adding more information until it is mitigated, may require a sequential approach \cite{chen2019continuous}.
\begin{figure}[h]
  \centering
   \includegraphics[width=0.8\linewidth]{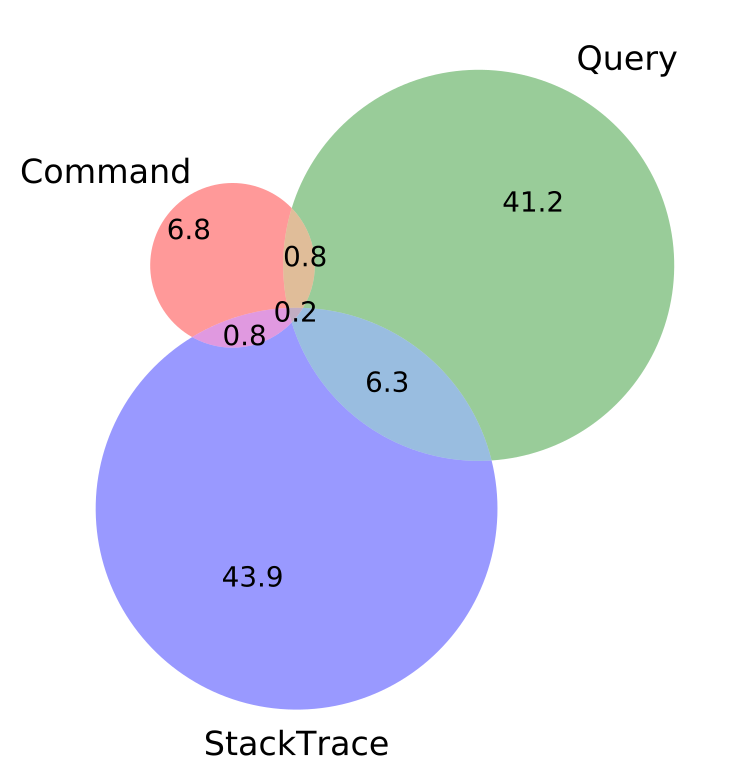}
  \caption{Venn diagram showing distribution of incidents containing command lines, queries, or stack traces (in percentage).}
  \Description{Data Type}
  \label{fig:dataType}
\end{figure}
\item The agile development model in cloud services require the development and deployment process being flexible and meeting certain production requirements in terms of execution time, or error tolerance to maintain high service availability. 
    \item Finally, introducing a new ML system into an existing workflow of engineers requires gaining their trust in adopting and getting feedback transfer actions.
\end{itemize}

\section{{\TransferAssistant} Overview} \label{sect:overview}
{\TransferAssistant} is an intelligent incident routing system built upon an ensemble of different machine learning techniques and runs on large scale online services. Besides combining the strength of state-of-the-art machine learning approaches, our system uses best practices in design and implementation to meet production requirements. Next, we present the end to end pipeline of {\TransferAssistant}. We start with ingesting data to a distributed database management system for efficient processing (data ingestion); cleaning and partitioning the data for large scale training (data processing); extracting features (feature engineering); and incrementally rolling out the system by adding more advanced machine learning models to address the feedback and improve performance from previous deployments (model architecture). 

\subsection{Data Ingestion}
{\TransferAssistant} uses resolved incidents in {\IcM} as training data. The label for each incident is the first team mitigating the incident and is \textit{automatically annotated} once a team mitigates an incident.

As the data is highly imbalanced (Figure \ref{fig:team_dist}), we merged all infrequent teams into a special team (class), called \textit{Other}. A low performance on the infrequent classes will not significantly affect the overall performance. 
Further, reducing the number of classes decreases model size and complexity, leading to lower training time and execution latency. 
We extract both textual and contextual data from the incident reports, which has been shown to be effective in prior work \cite{murphy2004automatic,chen2019continuous}. Textual data comes from the incident title, summary, and the first three discussion entries. Our analysis indicates that the important information for incident routing is covered in the first three entries and after that, most of the information is about understanding the root cause and logging the troubleshooting steps, as also observed by Chen et al. \cite{chen2019continuous}. 
Beside the textual features, we use other important contextual information from {\IcM} such as:
\begin{itemize}
    \item Features related to sources of an incident, e.g. source name (SourceName), the first assigned service (OriginatingServiceId), occurring device name (OccurringDeviceName), or raising data center (RaisingDC).
    \item Features related to characteristics of an incident, e.g. severity (Severity), incident type (IncidentType), or assigned keywords (Keyword).
\end{itemize}

Textual data, contextual data, and labels of incident reports are extracted from the SQL database of {\IcM} and ingested into Cosmos \cite{chaiken2008scope}, a distributed computing platform, for efficient batch processing. The ingestion process is done periodically which allows us retraining models with latest data within hours.

\subsection{Data Processing}
To be cost-effective and scale to large datasets, we partition the data into buckets and allow training models in parallel on cloud VMs with commodity CPUs. We sample at most $n$ data points from each class and put them into a bucket. The final model is an ensemble of models from different buckets. This approach addresses three problems in the original dataset: 1) reducing the training cost and time because the data size has been reduced significantly, 2) reduce data imbalance by down-sampling majority classes to at most $n$ instances, and 3) decrease the variance of predictions as this partitioning is the bagging method for ensemble \cite{jonsson2016automated}.

We further reduce noisy data by applying domain knowledge from {\IcM}. First, we sub-sample the data with more weights towards more recent data points. Because old data would be outdated as services continue to evolve, e.g. new teams onboarded and old teams deprecated.  Second, we set an upper limit of incidents having the same title and mitigated by the same team because many LSIs are generated from the same template and are mitigated by the same teams. Third, we discard most of the low severity incidents and make the majority (e.g., $80\%$) of the dataset of highly impacted incidents; note that most incidents are LSIs (Table \ref{tab:dataDist}) and our analysis revealed many of them are transient (do not need to be transferred to another team).

For textual data, we perform additional processing:
\begin{itemize}
    \item Cleaning HTML, XML tags, and binary data of embedded images
    \item Normalizing unicode characters and special entities, e.g. number, GUID, or URL
    \item Segmentation/tokenization: textual data is segmented into sentences. After that, each sentence is tokenized into individual words (tokens). We use the NLTK toolkit for text segmentation and tokenization.
\end{itemize}

\subsection{Feature Engineering}
With conventional machine learning models, we categorize contextual data and extract natural language processing (NLP) features for textual data:
\begin{itemize}
    \item Using n-gram hashing to handle the large vocabulary size from incident reports. Besides normal verbal discussion, incident reports have a large vocabulary because engineers put in other information, such as stack traces, error logs, or SQL queries.
    \item Removing noisy (repeated but not informative) phrases, e.g.  \textit{``Incident Created''} or \textit{``resource is unhealthy''}, and extracting key phrases. We use SysSeive toolkit \cite{potharaju2013juggling,potharajuconfseer} for key phrase extraction and removing the noisy phrases.
\end{itemize}
With deep learning models, we build domain-specific embeddings for both textual and contextual data.
\begin{itemize}
    \item Textual data: We build fastText embeddings \cite{bojanowski2017enriching} on the title, summary, and discussion.
    \item Contextual data: We concatenate all contextual features and build exponential family embeddings \cite{rudolph2016exponential}, i.e. an extension of textual word embeddings for high dimensional data.
\end{itemize}

\subsection{Model Architecture}
{\TransferAssistant} is an ensemble approach that combines both conventional machine learning and deep neural networks; ensemble is widely used in Kaggle and has been shown to be effective \cite{jonsson2016automated}. For a production environment, a key requirement is to take an agile approach, i.e., first roll out a base ML model to customers, get feedback, and then drive incremental improvement. We follow the same best practice to incrementally build  {\TransferAssistant}.
\subsubsection{Multiple Additive Regression Tree} In the first iteration, we built a simple multiple additive regression tree (MART) and partition into multiple models to scale up with the amount of data and classes (teams) in {\IcM}. We build a one-vs-all FastTree\cite{interlandi2019mlnet} binary classifier, a MART using gradient boosting algorithm in ML.NET. 
In each training step, FastTree builds a decision tree to add to the ensemble of trees from previous steps. The trained MART is the ensemble of these decision trees.

We observed the diminishing returns on using a large number of features for MART. Based on preliminary experiments on a validation dataset, \textit{MART} takes the top 30,000 features as its input using Mutual Information feature selection.

To scale up the system, we partition the data into 10 different buckets leading to 10 MART models. The final recommendations are the top 5 classes ensembled from these 10 models.
\subsubsection{Light Gradient Boosting Machine for CRIs} The second iteration focused on the sparseness and unstructured data in CRIs (as compared to LSIs)
by adding a one-vs-all LightGBM which was trained using only CRIs, called LGBM(CRI). LightGBM is an implementation of the gradient boosting machine algorithm. We select the top 50,000 features.

Since we are only training on CRIs, we partition \textit{LGBM(CRI)} into 3 buckets. \textit{LGBM(CRI)} ensembles predictions from these 3 models and outputs the top 5 predictions. 
\subsubsection{Inverted Index} The third iteration bridged the gap in the previous models where new teams or short CRIs usually have low performance because of little training data. To address this \textit{cold start} problem, we build a simple inverted index (\textit{InvertedIndex}) consisting of two inverted index tables based on IDF. One table, called \textit{local\_table}, extracts the top 200 words from each team. The other table, called \textit{global\_table}, uses the top 500 words from each team and is re-ranked by IDF scores. We extract unigrams from all textual data as input features. \textit{InvertedIndex} calculates the confidence score of each team as $confidence = \alpha(local\_table) + (1 - \alpha)(global\_table)$ where $\alpha=0.2$. The final recommendation list is the top 5 confidence scores.
\subsubsection{Content-based approach} In the fourth iteration, we employed a content-based approach to push improvements on the previous \textit{cold start} issue as well as on all incidents. We observed that similar incidents are likely to be mitigated by the same teams. We build a locality-sensitive hashing model (\textit{SI}), an unsupervised approach approximating nearest neighbor search algorithms, to efficiently find similar incidents for the current incident. Because \textit{SI} is an unsupervised approach, it can address the \textit{cold start} issue and scale well with minority classes.\\
In the inference stage, \textit{SI} loops through all incidents in the training data which is very expensive. We partition training data into 10 buckets so \textit{SI} can search all the buckets in parallel.
\subsubsection{Deep Neural Network} 
\begin{figure}[h]
  \centering
  \includegraphics[width=\linewidth]{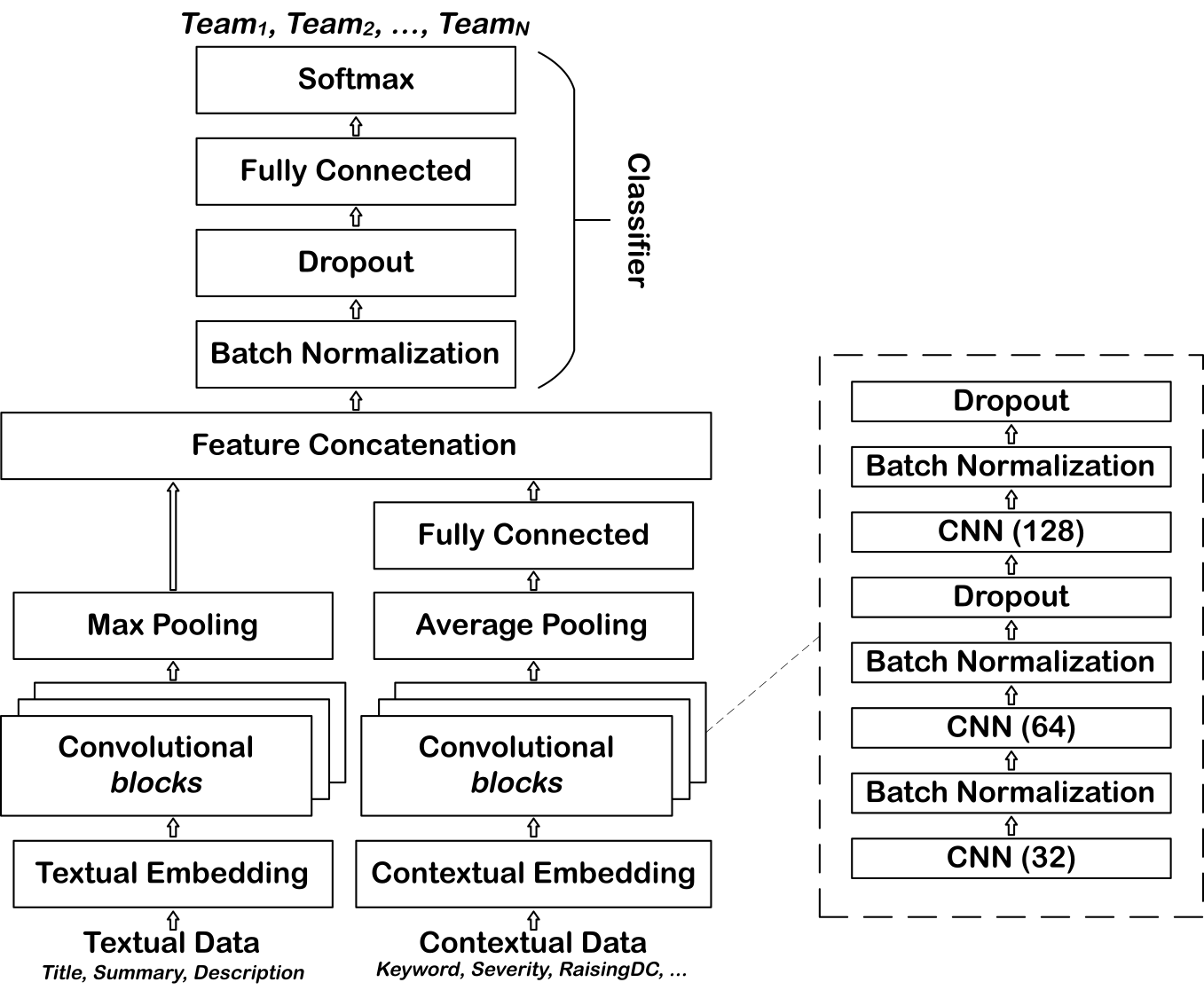}
  \caption{The Convolutional Neural Network architecture}
  \Description{CNN architecture}
  \label{fig:deepTriage}
\end{figure}

In the latest iteration, we use deep neural networks (\textit{DNN}) with domain-specific textual and contextual encoding. Similar to previous work \cite{chen2019continuous, lee2017applying}, we use the state-of-the-art Convolutional Neural Network (CNN) architecture for \textit{DNN} (Figure \ref{fig:deepTriage}). Both textual and contextual embeddings, extracted from an incident report, are fed through a \textit{convolutional block}. A convolutional block consists of CNN layers followed by batch normalization (BN) and scaled exponential linear unit (SELU) \cite{klambauer2017self} layers. We increase the number of filters after each CNN layer, i.e. 32, 64, and 128, to force the network gradually learning good representations of the data. To avoid overfitting, besides BN, we use dropout after the second and the third CNN layer. After the convolutional block, the contextual encoder uses an average pooling layer followed by a fully connected layer while the textual encoder uses a max pooling layer. Both textual and contextual representations are concatenated before forwarded to the classifier component. The classifier component consists of a BN, a dropout, a fully connected layer, and a softmax layer. \textit{DNN}'s output is the top 5 classes (teams) with the highest scores.
\subsubsection{Ensemble model (\textit{{\TransferAssistant}})} Taking all outputs from previous models (10 \textit{MART}s + 3 \textit{LGBM(CRI)}s + 1 \textit{InvertedIndex} + 1 \textit{SI} + 1 \textit{DNN} = 16 models), \textit{{\TransferAssistant}} re-ranks confidence scores of predictions and gives the top 5 recommendations. Different from a classification system, a recommendation system like {\TransferAssistant} is effective as long as the correct prediction is within the top $N$ ($N=5$) recommendations. Therefore, we maximize the $Recall@N$ rather than the $Accuracy@N$, i.e. keep the highest confidence score of each team across all models and select the top 5 teams.

\section{Evaluation} \label{sect:experiment}
\subsection{Settings \& Evaluation Metric}
Our experiment aims to answer three key questions:
\begin{itemize}
    \item \textbf{Q1}: Can {\TransferAssistant} give highly accurate recommendations to transfer incidents in {\IcM}?
    \item \textbf{Q2}: Does {\TransferAssistant} have good performance in highly impacted incidents e.g., high severity incidents, incidents in core services, CRIs?
    \item \textbf{Q3}: Does each incremental model provide a lift in performance for incident routing?
\end{itemize}
We collected six months of data for the training set and retrieved another three days of data for testing. We analyze the results in six scenarios: all incidents (\textit{All}), high severity incidents (\textit{Sev0-2}), \textit{difficult} incidents as having at least 2 hops (\textit{Min2Hop}), only customer reported incidents (\textit{CRI}), high severity incidents from the core services in Azure (\textit{Sev0-2@CoreServices}), and CRIs having high severity in Azure's core services, called \textit{CRI(Sev0-2)@CoreServices}.

The training set has more than 80GB of incident reports, collected from more than 4,000 teams and more than 1,000 services. The data is very imbalanced with the most popular teams having more than hundreds of thousands of incidents while minority teams have only one incident (Figure \ref{fig:team_dist}). Note that due to the policy of Microsoft, we cannot disclose the actual number of incident reports in this paper.

\begin{realnumber}
\begin{table}
  \caption{Distribution of incidents in the train set and test set.}
  \label{tab:dataDist}
  \begin{tabular}{|l|r|r|}
    \hline
    \textbf{Area}&\textbf{Train set}&\textbf{Test set}\\
    \hline
    Sev0-2&$8.23\%$&$7.07\%$\\
    \hline
    Min2Hop&$1.24\%$&$0.70\%$\\
    \hline
    CRI&$3.01\%$&$2.17\%$\\
    \hline
    Sev0-2@CoreServices&$0.26\%$&$0.33\%$\\
    \hline
    CRI(Sev0-2)@CoreServices&$0.03\%$&$0.03\%$\\
  \hline
   All&$100.00\%$&$100.00\%$\\
    \hline
\end{tabular}
\end{table}
\end{realnumber}

As {\TransferAssistant} returns an ordered list of 5 recommendations, we use the common performance metrics for recommendation systems or information retrieval, i.e. Precision@N, Recall@N, and F1@N where $N\in[1..5]$.

\subsection{Experiment Result}
\begin{figure*}[h]
  \centering
     \includegraphics[width=\linewidth]{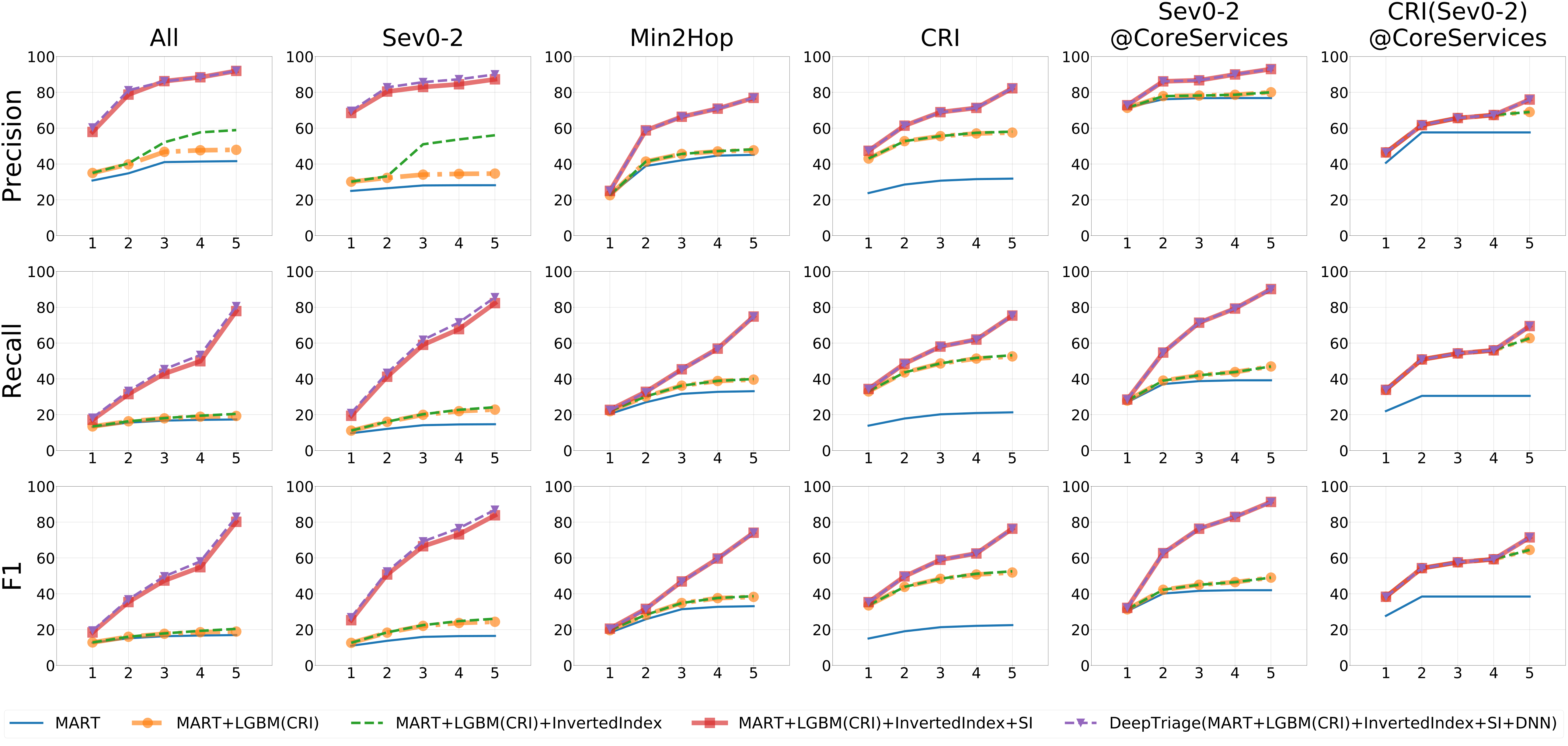}
  \caption{Experimental results: $F1=82.95\%$ on all incidents and $F1$ in range $71.49\%-91.15\%$ in highly impacted incidents. \textit{LGBM(CRI)} added $2\%$ to $29\%$ $F1$ improvement to scenarios related to CRIs upon \textit{MART}. Adding \textit{InvertedIndex} brought $11\%$ to $21\%$ $Precision$ improvement for all incidents or high severity incidents. \textit{SI} pushed improvements in all metrics, even with top 1 recommendation ($Precision@1$ had $+23\%$ in all incidents and $+38\%$ in high severity incidents). \textit{DNN} lifted $F1$ in all incidents ($+2.73\%$) and in high severity incidents ($+3.04\%$) as well as top 1 recommendations, e.g. in all incidents $Precision@1 +2.41\%, Recall@1 +1.12\%$.}
  \Description{Performance Metrics}
  \label{fig:perf}
\end{figure*}

Figure \ref{fig:perf} shows the performance@N of {\TransferAssistant} in each evaluating scenario. In general, {\TransferAssistant} achieved a good performance ($F1=82.95\%$) on all incidents (\textbf{Q1}) and highly impacted scenarios (\textbf{Q2}): \textit{Sev0-2} ($F1=86.88\%$), \textit{CRI} ($F1=76.38\%$), \textit{Sev0-2@CoreServices} ($F1=91.15\%$), and \textit{CRI(Sev0-2)@CoreServices} ($F1=71.49\%$). 
The system struggled with \textit{difficult} incidents ($F1=74.14\%$ in \textit{Min2Hop}). 
Further investigations show that most of the long-hop incidents require multiple teams engage to analyze and find the root cause.
For example, a lot of disconnected virtual machines could be due to network device failures or power loss or usage throttling. Finding the right team in these incidents requires additional debugging data.\\
Last but not least, the final model got modest performance at top 1 recommendations. This is due to the current ensemble policy, where {\TransferAssistant} is optimized for team coverage but not accuracy.

To answer \textbf{Q3}, we analyze the model ablation results to find the benefit of adding each model component to {\TransferAssistant}.\\
First, combining \textit{LGBM(CRI)} with \textit{MART} gave $F1$ improvements in all evaluated scenarios, especially in \textit{CRI}. The improvements in $F1$, ranging from $+2\%$ (\textit{All}) to $+29\%$ (\textit{CRI}), is credited mostly to the improvements in $Recall$.
Moreover, \textit{Sev0-2} also benefited from the \textit{MART+LGBM(CRI)} model in all performance metrics compared to \textit{All}. We think this improvement comes from the increasing ratio of CRIs in \textit{Sev0-2} ($3.55\%$) compared to \textit{All} ($2.24\%$). These results suggested that using LGBM(CRI) models, trained specifically on CRI data, will help with routing more CRIs to the right teams.

Second, adding \textit{InvertedIndex} upon \textit{MART+LGBM(CRI)} did not give a significant improvement for {\TransferAssistant}. However, we observed some improvements in $Precision$ for incidents in \textit{All} ($+11\%$) and \textit{Sev0-2} ($21\%$) but not much in $Recall$. This result is expected as \textit{InvertedIndex} targets the $cold start$ situation where the incidents do not frequently appear. Therefore, the $Recall$ metrics did not benefit much from this approach.

Third, adding \textit{SI} gave a significant improvement for {\TransferAssistant} in all performance metrics. The improvements can be seen clearly in $Precision@1$ for incidents in \textit{All} ($+23\%$) and \textit{Sev0-2} ($+38\%$). This result implies that many incidents from our data are similar (LSIs generated from templates and mitigated within the same team). The big improvements from $Recall@4$ and $Recall@5$ for CRIs in \textit{CRI} and \textit{CRI(Sev0-2)@CoreServices} also implied a counter intuitive assumption, i.e. human-generated incidents are very similar. We analyzed these cases further and found those teams use predefined templates for incidents created toward them. These templates ask users to input required data before transferring the ticket to avoid extra transfers because of missing important information.

Fourth, \textit{DNN} slightly improved {\TransferAssistant} in \textit{All} ($+2.73\%$ in $F1$) and \textit{Sev0-2} ($+3.04\%$ in $F1$).
Interestingly, we observed that \textit{DNN} gave improvements at top 1 performance from both $Precision$ ($+2.41\%$ in \textit{All} and $+1.14\%$ in \textit{Sev0-2}) and $Recall$ ($+1.12\%$ in \textit{All} and $+1.52\%$ in \textit{Sev0-2}). This top 1 improvement suggested that \textit{DNN} will bring more value to {\TransferAssistant} when we deploy the system as a fully automated routing service.

\section{Production System} \label{sect:production}
{\TransferAssistant} has three main components:
\begin{itemize}
    \item Controller: is implemented as a part of the {\IcM} code base. This component handles HTTP requests, collects incident related data, coordinates the other 2 components, and returns top 5 recommendations for each requested incident. 
    \item Predicting: ensembles 16 ML models: 10 \textit{MARTs}, 3 \textit{LGBM(CRI)s}, 1 \textit{SI}, 1 \textit{InvertedIndex}, 1 \textit{SI}, and 1 \textit{DNN}; which are deployed as Azure Machine Learning (AML) web services. Each web service has a data preprocessing module and a machine learning model which outputs 5 recommended teams and their confidence scores.
    \item Ensemble: is implemented as an integrated part of the {\IcM} code base. The Ensemble component takes outputs from all models, applies the re-rank algorithm, and returns the top 5 recommendations (teams) to the Controller component. 
\end{itemize}

\subsection{{\TransferAssistant} Usage in Azure}
Implemented as a web service makes it easy to apply {\TransferAssistant} into multiple use cases. Our strategy is rolling {\TransferAssistant} out as an on-demand assisting service and gradually move humans out of the loop to create a fully automated incident routing service for {\IcM}. Next, we describe 4 different use cases that we have been working on to implement the road map.

\textit{API call}: customers use an HTTP Client to call the {\TransferAssistant} API. 
Customers pass an incident ID and receives the top 5 recommended teams that the incident should be transferred to, an auto-suggest. This use case fits the initial stage where we can test the system functionality and allow pilot customers evaluate the correctness of the system without any UI support.

\textit{On the incident's detail web page}: when an engineer has problems with selecting a team to transfer an incident to. The engineer can click the Transfer button on {\IcM} portal. {\TransferAssistant} will collect the incident ID from the current {\IcM} web page, process, and  return the 5 most potential teams for the requested incident.

\textit{At creation time}: engineers can also take advantage of {\TransferAssistant} right at the time they create incidents. Different from the previous use cases, the incident does not have an incident ID because, literally, it is not created yet. We modify the Controller component to receive title and summary of the incident being created and get top 5 recommendations from {\TransferAssistant}.

\textit{Automation workflow}: we use the LogicApps (LA) system, an automated workflow based on Azure Logic Apps \cite{logicapps}, to introduce an initial experience of automated incident routing. We create a workflow listening to all incidents coming to a team. When an incident arrives to the team, the workflow can get 5 recommendations for the incident by calling {\TransferAssistant}. The key difference in this use case is the workflow needs to select 1 from the 5 recommendations to transfer the incident. While the top 1 performance of {\TransferAssistant} is not perfect, we can improve the performance by incorporating domain knowledge as logic rules in the workflow. 
We have received a large quantity of positive feedback and observed good results from this kind of machine learning and domain knowledge combination.
\subsection{Scaling \& Reliability}
\subsubsection{Scaling} We have applied best practices in multiple aspects of {\TransferAssistant} to scale up in both training (Section \ref{sect:overview}) and execution. In model design, we partition data to create multiple models trained and executed in parallel. In model building, we use Azure Batch \cite{azurebatch} to acquire VMs on demand and train models in parallel. In model deployment, we deployed the models as AML web services which can be scaled on demand.

Deployed since 2017 on {\IcM}, {\TransferAssistant} serves approximately $1,000$ unique customers with more than $1,500$ API calls per day. 
 The API latency is approximate $833.6 ms$.

\subsubsection{Reliability}
Keeping the system's availability high is the key point to support incident triage. As {\TransferAssistant} is an ensemble of 16 AML web services and the ranking component is hosted on the {\IcM} system, we have dependencies on these two sources. With AML web services, we observed that most of the errors from AML web services are self-mitigated within a short time and all 16 models have never been down at the same time. We address this problem with a fail-over mechanism. {\TransferAssistant} only returns a failure status when all 16 models cannot return any recommendations. Otherwise, the \textit{ensemble} component will use all available results to pick the top 5 recommendations. 
A drawback of this mechanism is {\TransferAssistant} will miss the correct team when all models suggesting this team all fail. 
However, this is a reasonable trade off because the chance that many models fail at the same time is very low. Besides, the Controller and the Ensemble components inherit the redundancy mechanism from {\IcM}. These components are run on two independent platforms and only fail when both platforms are down.

Another reliability problem is data distribution shift: the model's accuracy decreases when data distribution shifts along the time.
We setup a retraining pipeline to refresh the models periodically. Currently, we use the last 3 day as testing data. To improve the quality of the test set and make the process more data-driven, we are looking at machine learning approaches as in \cite{chung2019slice} to scale up and automate the validation step.

\section{Lessons Learned} \label{sect:discussion}
\textit{Different customers find {\TransferAssistant} helpful in different ways}.
After deploying {\TransferAssistant}, we collected customer feedback for the service.

Junior engineers found {\TransferAssistant} very helpful as they do not have enough experience about teams within the system. For example, ``\textit{It's really amazing. It helps us to transfer the issue to the right team faster than earlier}'' or ``\textit{in general this feature has been useful}''.

On the other hand, we did not receive as much positive feedback from senior engineers while the usage from these engineers is as high as the junior engineers. 
For example, ``\textit{I knew exactly where I wanted to transfer the incident when I opened it.}'' or ``\textit{it prompts the teams most suitable but it so happens that teams get re-org so often that scope changes from time to time and looking for the right team becomes a challenge}''. Further investigations showed that senior engineers used {\TransferAssistant} as a ``\textit{shortcut}'' to transfer incidents to those teams. Because it takes a single click to transfer an incident with {\TransferAssistant} when the correct team is in the recommendation list while it takes more efforts to navigate to the same team from the current UI.

\textit{Provide Insights behind ML recommendations to gain user trust}. One of the biggest problems we have when deploying {\TransferAssistant} is gaining the trust from our customers, i.e. OCEs, especially the senior engineers who have a lot of domain knowledge and have doubts about the models' suggestions.
To quickly address this challenge, we developed a new feature for {\TransferAssistant}, called \textit{RecommendationInsights}, providing additional information about the models' suggestions to strengthen the trust from OCEs. For each recommendation, \textit{RecommendationInsights} provides how many incidents, which are similar to the current incident, were mitigated in the same team as this recommendation, by how many unique OCEs, and the list of keywords from these similar incidents. This additional information gives OCEs more confidence that they are doing the right thing with {\TransferAssistant}.

\textit{Handle every case, no corner cases!} One of the main problems with supervised learning is annotated data. While the incident triage problem is self-annotated, we still need to solve the ``cold start'' problem, where new teams do not have sufficient data. Therefore, {\TransferAssistant} includes the \textit{InvertedIndex} and \textit{SI} models which are unsupervised approaches. The new teams would rely more on similar incident models until they have sufficient training data.

\textit{A perfect {\TransferAssistant} cannot completely drive down transfer hops}. The number of hops in some teams will not reduce even with an oracle system. The internal structure of some services requires some \textit{transfer loops} for communication across the service boundary. For example, internal teams do not have direct connects with customers and need to use another customer-facing team as a communication protocol. Therefore, even if {\TransferAssistant} suggests the correct internal team, the incident still takes the \textit{transfer loops} until it is mitigated (by the recommended team!).

\textit{Identify and address label gaps}. From error analysis, we found out some reasons for error predictions in {\TransferAssistant}. First, there is a gap in labeling. By internal policy, some teams mitigate incidents even when they are not the teams who are responsible for the incidents.
To address this problem, we introduce a new UI component on {\IcM} portal, called ResponsibleTeamId, for each incident. With the new workflow, teams are able to mitigate incidents not belonging to them as long as they set the correct mitigating team for the ResponsibleTeamId.

Second, some incidents do not have sufficient information for triage. When receiving an incident, on-call engineers would join a call bridge or run some validation scripts on their separated systems to find the root cause and the right triage teams. While the bridge discussions are in {\IcM} and we can incorporate into {\TransferAssistant}, the debugging systems are not in {\IcM} yet. We have started working with teams and services to enrich incidents with this information, e.g. using automation workflows to add these information to incidents automatically.

\textit{Combine the best of both worlds, machine learning and domain knowledge}. The semi-automated incident triage (LA Workflow) is introduced, teams use their domain knowledge to improve the top 1 performance of {\TransferAssistant}.

The top recommendation from {\TransferAssistant} is used when none of the rules can handle an incident. This approach actually boosted the performance of the workflow above $90\%$ accuracy as the rules have very high precision. We think it is a pragmatic way to engage teams to use {\TransferAssistant} as an automated service rather than manually triggering. As many teams build their own rules combining with {\TransferAssistant}, we can collect those rules and enrich {\TransferAssistant}'s features to capture all the domain knowledge. However, there is a drawback with this approach as teams need to manage the workflow by themselves. {\TransferAssistant} can be used as a fully automated service to route incidents out of a team if the service has a high confidence that the incident should be in another team. We have a plan to develop this fully automated service. One of the key challenges of this is improving the \textit{performance@1}. This can be augmented by using domain knowledge from current LA workflows.
\section{Related Work} \label{sect:relatedWork}
Bug triage is very close to the problem we are solving where the system gives a ranked list of developers (solvers) for a bug report. Early work in bug triage using textual data from the report's description \cite{murphy2004automatic} and contextual features, e.g. product related data, from both the current report and previous reports \cite{xia2013accurate}. While this approach has suffered from the sparseness between bug reports and new developers, researchers have leveraged developer's expertise to increase the system's accuracy. Matter et al. \cite{matter2009assigning} extract a developer's expertise from all codes that the developer contributed. On the other hand, Xie et al. \cite{xie2012dretom} used topic modeling (Latent Dirichlet Allocation) to calculate the probability assigning a developer to the most dominant topic of a bug report.

In more specific context, researchers tried to reduce number of reassignments (bug tossing) with Markov-based models, called bug tossing graph, from textual data \cite{jeong2009improving} with contextual (product-related) data \cite{bhattacharya2012automated}. In addition to a single approach, Jonsson et al. \cite{jonsson2016automated} showed ensembling multiple models, e.g. Naive Bayes, SVM, KNN, and Decision Tree, will increase the accuracy.

Recently, with the emergence of deep neural networks, researchers have explored Convolutional Neural Network \cite{lee2017applying} with textual data (description and discussion) and sequence to sequence model \cite{xi2019bug} on textual content and tossing sequence. Moreover, \cite{lee2017applying} also reported their experience when deploying the approach in an industrial environment: using a human assistant instead of fully automated and the need to continuous refreshing (re-training) the model. 

Besides all of the problems managing incidents (bugs) submitted by users, an online service's management system also proactively detects issues from its components using telemetry data (logs) \cite{xu2009detecting, cohen2004correlating, cohen2005capturing, bodik2010fingerprinting}. 

Going beyond incident detection, Lou et al. \cite{lou2013software}  tried to address the ``fixing'' stage by analyzing similar incidents and extracting actions to mitigate the current incidents. In additional to logs, an incident management system also handles CRIs \cite{lou2017experience}.

Recently, DeepCT \cite{chen2019continuous} proposed a Gated Recurrent Unit (GRU) model treating description entries (from both LSIs and CRIs) as a sequence of signal to improve the incident triage process. 

{\TransferAssistant} took the best practices from previous work in bug triage and incident triage, i.e. combining textual and contextual data, ensemble multiple approaches including deep neural network, and frequently retraining. While addressing the same incident triage problem, there are major differences between \cite{chen2019continuous} and our work: we do not only focus on deep neural network but ensemble it with other machine learning approaches and we shared lessons learned when deploying {\TransferAssistant} in a production environment. Lee et al. \cite{lee2017applying} also reported a deployed bug triage system. However, they aimed to assist human to triage bugs, while we work on online services and follow a road map from assisting human to semi-automated and, eventually, fully-automated triage incidents.

\section{Conclusion \& Future work} \label{sect:conclusion}
In this paper, we introduced {\TransferAssistant}, a novel machine learning service that improves the incident triage process by providing recommendations for the right team to mitigate an incident. The system has been deployed on an incident management system for Azure since 2017. We described how {\TransferAssistant} was implemented, scaled up, and how it assists on-call engineers at multiple levels from manual to automated. We did performance analysis on the ensemble model of the system and showed how the final prediction incrementally improved from its sub-models. Finally, we share the lessons learned in the development and deployment of {\TransferAssistant}.

In the future, we plan to 
introduce an auto routing service which automatically tracks incidents assigned to teams. If the auto routing service has a high confidence that the incident belongs to a different responsible team, it will transfer the incident to the correct team automatically. 
However, a fundamental challenge for this task is the \textit{performance@1}. As a recommendation system, it does not impact users much when the correct team is one of the top 5 recommendations. However, for auto-routing, we need to improve \textit{performance@1} because no human will intervene and correct the choice for {\TransferAssistant}.
Further, we will improve the \textit{DNN} model of {\TransferAssistant} and experiment more state-of-the-art techniques from Deep Neural Networks such as Transformers to improve the accuracy of {\TransferAssistant}.

\bibliographystyle{ACM-Reference-Format}
\bibliography{transferAssistant_kdd2020}


\begin{thebibliography}{33}


\ifx \showCODEN    \undefined \def \showCODEN     #1{\unskip}     \fi
\ifx \showDOI      \undefined \def \showDOI       #1{#1}\fi
\ifx \showISBNx    \undefined \def \showISBNx     #1{\unskip}     \fi
\ifx \showISBNxiii \undefined \def \showISBNxiii  #1{\unskip}     \fi
\ifx \showISSN     \undefined \def \showISSN      #1{\unskip}     \fi
\ifx \showLCCN     \undefined \def \showLCCN      #1{\unskip}     \fi
\ifx \shownote     \undefined \def \shownote      #1{#1}          \fi
\ifx \showarticletitle \undefined \def \showarticletitle #1{#1}   \fi
\ifx \showURL      \undefined \def \showURL       {\relax}        \fi
\providecommand\bibfield[2]{#2}
\providecommand\bibinfo[2]{#2}
\providecommand\natexlab[1]{#1}
\providecommand\showeprint[2][]{arXiv:#2}

\bibitem[\protect\citeauthoryear{Ahmed, Amizadeh, Bilenko, Carr, Chin, Dekel,
  Dupre, Eksarevskiy, Filipi, Finley, Goswami, Hoover, Inglis, Interlandi,
  Kazmi, Krivosheev, Luferenko, Matantsev, Matusevych, Moradi, Nazirov, Ormont,
  Oshri, Pagnoni, Parmar, Roy, Siddiqui, Weimer, Zahirazami, and Zhu}{Ahmed
  et~al\mbox{.}}{2019}]%
        {interlandi2019mlnet}
\bibfield{author}{\bibinfo{person}{Zeeshan Ahmed}, \bibinfo{person}{Saeed
  Amizadeh}, \bibinfo{person}{Mikhail Bilenko}, \bibinfo{person}{Rogan Carr},
  \bibinfo{person}{Wei-Sheng Chin}, \bibinfo{person}{Yael Dekel},
  \bibinfo{person}{Xavier Dupre}, \bibinfo{person}{Vadim Eksarevskiy},
  \bibinfo{person}{Senja Filipi}, \bibinfo{person}{Tom Finley},
  \bibinfo{person}{Abhishek Goswami}, \bibinfo{person}{Monte Hoover},
  \bibinfo{person}{Scott Inglis}, \bibinfo{person}{Matteo Interlandi},
  \bibinfo{person}{Najeeb Kazmi}, \bibinfo{person}{Gleb Krivosheev},
  \bibinfo{person}{Pete Luferenko}, \bibinfo{person}{Ivan Matantsev},
  \bibinfo{person}{Sergiy Matusevych}, \bibinfo{person}{Shahab Moradi},
  \bibinfo{person}{Gani Nazirov}, \bibinfo{person}{Justin Ormont},
  \bibinfo{person}{Gal Oshri}, \bibinfo{person}{Artidoro Pagnoni},
  \bibinfo{person}{Jignesh Parmar}, \bibinfo{person}{Prabhat Roy},
  \bibinfo{person}{Mohammad~Zeeshan Siddiqui}, \bibinfo{person}{Markus Weimer},
  \bibinfo{person}{Shauheen Zahirazami}, {and} \bibinfo{person}{Yiwen Zhu}.}
  \bibinfo{year}{2019}\natexlab{}.
\newblock \showarticletitle{Machine Learning at Microsoft with ML.NET}. In
  \bibinfo{booktitle}{\emph{Proceedings of the 25th ACM SIGKDD International
  Conference on Knowledge Discovery \& Data Mining}} (Anchorage, AK, USA)
  \emph{(\bibinfo{series}{KDD ’19})}. \bibinfo{publisher}{Association for
  Computing Machinery}, \bibinfo{address}{New York, NY, USA},
  \bibinfo{pages}{2448–2458}.
\newblock
\showISBNx{9781450362016}
\urldef\tempurl%
\url{https://doi.org/10.1145/3292500.3330667}
\showDOI{\tempurl}


\bibitem[\protect\citeauthoryear{AppDynamics}{AppDynamics}{2015}]%
        {appdynamics}
\bibfield{author}{\bibinfo{person}{AppDynamics}.}
  \bibinfo{year}{2015}\natexlab{}.
\newblock \bibinfo{booktitle}{\emph{The Real Cost of Downtime, The Real
  Potential of DevOps}}.
\newblock
\urldef\tempurl%
\url{https://www.appdynamics.com/blog/engineering/idc-devops-cost-downtime}
\showURL{%
Retrieved February 13, 2020 from \tempurl}


\bibitem[\protect\citeauthoryear{Azure}{Azure}{2020a}]%
        {azurebatch}
\bibfield{author}{\bibinfo{person}{Microsoft Azure}.}
  \bibinfo{year}{2020}\natexlab{a}.
\newblock \bibinfo{booktitle}{\emph{Batch: Cloud-scale job scheduling and
  compute management}}.
\newblock
\urldef\tempurl%
\url{https://azure.microsoft.com/en-us/services/batch/}
\showURL{%
Retrieved February 13, 2020 from \tempurl}


\bibitem[\protect\citeauthoryear{Azure}{Azure}{2020b}]%
        {logicapps}
\bibfield{author}{\bibinfo{person}{Microsoft Azure}.}
  \bibinfo{year}{2020}\natexlab{b}.
\newblock \bibinfo{booktitle}{\emph{Logic Apps: Quickly build powerful
  integration solutions}}.
\newblock
\urldef\tempurl%
\url{https://azure.microsoft.com/en-us/services/logic-apps/}
\showURL{%
Retrieved February 13, 2020 from \tempurl}


\bibitem[\protect\citeauthoryear{Bhattacharya, Neamtiu, and
  Shelton}{Bhattacharya et~al\mbox{.}}{2012}]%
        {bhattacharya2012automated}
\bibfield{author}{\bibinfo{person}{Pamela Bhattacharya},
  \bibinfo{person}{Iulian Neamtiu}, {and} \bibinfo{person}{Christian~R
  Shelton}.} \bibinfo{year}{2012}\natexlab{}.
\newblock \showarticletitle{Automated, highly-accurate, bug assignment using
  machine learning and tossing graphs}.
\newblock \bibinfo{journal}{\emph{Journal of Systems and Software}}
  \bibinfo{volume}{85}, \bibinfo{number}{10} (\bibinfo{year}{2012}),
  \bibinfo{pages}{2275--2292}.
\newblock


\bibitem[\protect\citeauthoryear{Bodik, Goldszmidt, Fox, Woodard, and
  Andersen}{Bodik et~al\mbox{.}}{2010}]%
        {bodik2010fingerprinting}
\bibfield{author}{\bibinfo{person}{Peter Bodik}, \bibinfo{person}{Moises
  Goldszmidt}, \bibinfo{person}{Armando Fox}, \bibinfo{person}{Dawn~B Woodard},
  {and} \bibinfo{person}{Hans Andersen}.} \bibinfo{year}{2010}\natexlab{}.
\newblock \showarticletitle{Fingerprinting the datacenter: automated
  classification of performance crises}. In
  \bibinfo{booktitle}{\emph{Proceedings of the 5th European conference on
  Computer systems}}. \bibinfo{pages}{111--124}.
\newblock


\bibitem[\protect\citeauthoryear{Bojanowski, Grave, Joulin, and
  Mikolov}{Bojanowski et~al\mbox{.}}{2017}]%
        {bojanowski2017enriching}
\bibfield{author}{\bibinfo{person}{Piotr Bojanowski}, \bibinfo{person}{Edouard
  Grave}, \bibinfo{person}{Armand Joulin}, {and} \bibinfo{person}{Tomas
  Mikolov}.} \bibinfo{year}{2017}\natexlab{}.
\newblock \showarticletitle{Enriching word vectors with subword information}.
\newblock \bibinfo{journal}{\emph{Transactions of the Association for
  Computational Linguistics}}  \bibinfo{volume}{5} (\bibinfo{year}{2017}),
  \bibinfo{pages}{135--146}.
\newblock


\bibitem[\protect\citeauthoryear{Chaiken, Jenkins, Larson, Ramsey, Shakib,
  Weaver, and Zhou}{Chaiken et~al\mbox{.}}{2008}]%
        {chaiken2008scope}
\bibfield{author}{\bibinfo{person}{Ronnie Chaiken}, \bibinfo{person}{Bob
  Jenkins}, \bibinfo{person}{Per-{\AA}ke Larson}, \bibinfo{person}{Bill
  Ramsey}, \bibinfo{person}{Darren Shakib}, \bibinfo{person}{Simon Weaver},
  {and} \bibinfo{person}{Jingren Zhou}.} \bibinfo{year}{2008}\natexlab{}.
\newblock \showarticletitle{SCOPE: easy and efficient parallel processing of
  massive data sets}.
\newblock \bibinfo{journal}{\emph{Proceedings of the VLDB Endowment}}
  \bibinfo{volume}{1}, \bibinfo{number}{2} (\bibinfo{year}{2008}),
  \bibinfo{pages}{1265--1276}.
\newblock


\bibitem[\protect\citeauthoryear{Chen, He, Lin, Xu, Zhang, Hao, Gao, Xu, Dang,
  and Zhang}{Chen et~al\mbox{.}}{2019a}]%
        {chen2019empirical}
\bibfield{author}{\bibinfo{person}{Junjie Chen}, \bibinfo{person}{Xiaoting He},
  \bibinfo{person}{Qingwei Lin}, \bibinfo{person}{Yong Xu},
  \bibinfo{person}{Hongyu Zhang}, \bibinfo{person}{Dan Hao},
  \bibinfo{person}{Feng Gao}, \bibinfo{person}{Zhangwei Xu},
  \bibinfo{person}{Yingnong Dang}, {and} \bibinfo{person}{Dongmei Zhang}.}
  \bibinfo{year}{2019}\natexlab{a}.
\newblock \showarticletitle{An empirical investigation of incident triage for
  online service systems}. In \bibinfo{booktitle}{\emph{2019 IEEE/ACM 41st
  International Conference on Software Engineering: Software Engineering in
  Practice (ICSE-SEIP)}}. IEEE, \bibinfo{pages}{111--120}.
\newblock


\bibitem[\protect\citeauthoryear{Chen, He, Lin, Zhang, Hao, Gao, Xu, Dang, and
  Zhang}{Chen et~al\mbox{.}}{2019b}]%
        {chen2019continuous}
\bibfield{author}{\bibinfo{person}{Junjie Chen}, \bibinfo{person}{Xiaoting He},
  \bibinfo{person}{Qingwei Lin}, \bibinfo{person}{Hongyu Zhang},
  \bibinfo{person}{Dan Hao}, \bibinfo{person}{Feng Gao},
  \bibinfo{person}{Zhangwei Xu}, \bibinfo{person}{Yingnong Dang}, {and}
  \bibinfo{person}{Dongmei Zhang}.} \bibinfo{year}{2019}\natexlab{b}.
\newblock \showarticletitle{Continuous incident triage for large-scale online
  service systems}. In \bibinfo{booktitle}{\emph{2019 34th IEEE/ACM
  International Conference on Automated Software Engineering (ASE)}}. IEEE,
  \bibinfo{pages}{364--375}.
\newblock


\bibitem[\protect\citeauthoryear{Choi, Bahadori, Sun, Kulas, Schuetz, and
  Stewart}{Choi et~al\mbox{.}}{2016}]%
        {choi2016retain}
\bibfield{author}{\bibinfo{person}{Edward Choi}, \bibinfo{person}{Mohammad~Taha
  Bahadori}, \bibinfo{person}{Jimeng Sun}, \bibinfo{person}{Joshua Kulas},
  \bibinfo{person}{Andy Schuetz}, {and} \bibinfo{person}{Walter Stewart}.}
  \bibinfo{year}{2016}\natexlab{}.
\newblock \showarticletitle{Retain: An interpretable predictive model for
  healthcare using reverse time attention mechanism}. In
  \bibinfo{booktitle}{\emph{Advances in Neural Information Processing
  Systems}}. \bibinfo{pages}{3504--3512}.
\newblock


\bibitem[\protect\citeauthoryear{Chung, Kraska, Polyzotis, Tae, and
  Whang}{Chung et~al\mbox{.}}{2019}]%
        {chung2019slice}
\bibfield{author}{\bibinfo{person}{Yeounoh Chung}, \bibinfo{person}{Tim
  Kraska}, \bibinfo{person}{Neoklis Polyzotis}, \bibinfo{person}{Ki~Hyun Tae},
  {and} \bibinfo{person}{Steven~Euijong Whang}.}
  \bibinfo{year}{2019}\natexlab{}.
\newblock \showarticletitle{Slice finder: Automated data slicing for model
  validation}. In \bibinfo{booktitle}{\emph{2019 IEEE 35th International
  Conference on Data Engineering (ICDE)}}. IEEE, \bibinfo{pages}{1550--1553}.
\newblock


\bibitem[\protect\citeauthoryear{Cohen, Chase, Goldszmidt, Kelly, and
  Symons}{Cohen et~al\mbox{.}}{2004}]%
        {cohen2004correlating}
\bibfield{author}{\bibinfo{person}{Ira Cohen}, \bibinfo{person}{Jeffrey~S
  Chase}, \bibinfo{person}{Moises Goldszmidt}, \bibinfo{person}{Terence Kelly},
  {and} \bibinfo{person}{Julie Symons}.} \bibinfo{year}{2004}\natexlab{}.
\newblock \showarticletitle{Correlating Instrumentation Data to System States:
  A Building Block for Automated Diagnosis and Control.}. In
  \bibinfo{booktitle}{\emph{OSDI}}, Vol.~\bibinfo{volume}{4}.
  \bibinfo{pages}{16--16}.
\newblock


\bibitem[\protect\citeauthoryear{Cohen, Zhang, Goldszmidt, Symons, Kelly, and
  Fox}{Cohen et~al\mbox{.}}{2005}]%
        {cohen2005capturing}
\bibfield{author}{\bibinfo{person}{Ira Cohen}, \bibinfo{person}{Steve Zhang},
  \bibinfo{person}{Moises Goldszmidt}, \bibinfo{person}{Julie Symons},
  \bibinfo{person}{Terence Kelly}, {and} \bibinfo{person}{Armando Fox}.}
  \bibinfo{year}{2005}\natexlab{}.
\newblock \showarticletitle{Capturing, indexing, clustering, and retrieving
  system history}.
\newblock \bibinfo{journal}{\emph{ACM SIGOPS Operating Systems Review}}
  \bibinfo{volume}{39}, \bibinfo{number}{5} (\bibinfo{year}{2005}),
  \bibinfo{pages}{105--118}.
\newblock


\bibitem[\protect\citeauthoryear{Forbes}{Forbes}{2019}]%
        {gartner2022}
\bibfield{author}{\bibinfo{person}{Forbes}.} \bibinfo{year}{2019}\natexlab{}.
\newblock \bibinfo{booktitle}{\emph{Public Cloud Soaring To \$331B By 2022
  According To Gartner}}.
\newblock
\urldef\tempurl%
\url{https://www.forbes.com/sites/louiscolumbus/2019/04/07/public-cloud-soaring-to-331b-by-2022-according-to-gartner}
\showURL{%
Retrieved February 13, 2020 from \tempurl}


\bibitem[\protect\citeauthoryear{Insider}{Insider}{2018}]%
        {prime}
\bibfield{author}{\bibinfo{person}{Business Insider}.}
  \bibinfo{year}{2018}\natexlab{}.
\newblock \bibinfo{booktitle}{\emph{Amazon's one hour of downtime on Prime Day
  may have cost it up to \$100 million in lost sales}}.
\newblock
\urldef\tempurl%
\url{https://www.businessinsider.com/amazon-prime-day-website-issues-cost-it-millions-in-lost-sales-2018-7}
\showURL{%
Retrieved February 13, 2020 from \tempurl}


\bibitem[\protect\citeauthoryear{Jeong, Kim, and Zimmermann}{Jeong
  et~al\mbox{.}}{2009}]%
        {jeong2009improving}
\bibfield{author}{\bibinfo{person}{Gaeul Jeong}, \bibinfo{person}{Sunghun Kim},
  {and} \bibinfo{person}{Thomas Zimmermann}.} \bibinfo{year}{2009}\natexlab{}.
\newblock \showarticletitle{Improving bug triage with bug tossing graphs}. In
  \bibinfo{booktitle}{\emph{Proceedings of the 7th joint meeting of the
  European software engineering conference and the ACM SIGSOFT symposium on The
  foundations of software engineering}}. \bibinfo{pages}{111--120}.
\newblock


\bibitem[\protect\citeauthoryear{Jonsson, Borg, Broman, Sandahl, Eldh, and
  Runeson}{Jonsson et~al\mbox{.}}{2016}]%
        {jonsson2016automated}
\bibfield{author}{\bibinfo{person}{Leif Jonsson}, \bibinfo{person}{Markus
  Borg}, \bibinfo{person}{David Broman}, \bibinfo{person}{Kristian Sandahl},
  \bibinfo{person}{Sigrid Eldh}, {and} \bibinfo{person}{Per Runeson}.}
  \bibinfo{year}{2016}\natexlab{}.
\newblock \showarticletitle{Automated bug assignment: Ensemble-based machine
  learning in large scale industrial contexts}.
\newblock \bibinfo{journal}{\emph{Empirical Software Engineering}}
  \bibinfo{volume}{21}, \bibinfo{number}{4} (\bibinfo{year}{2016}),
  \bibinfo{pages}{1533--1578}.
\newblock


\bibitem[\protect\citeauthoryear{Klambauer, Unterthiner, Mayr, and
  Hochreiter}{Klambauer et~al\mbox{.}}{2017}]%
        {klambauer2017self}
\bibfield{author}{\bibinfo{person}{G{\"u}nter Klambauer},
  \bibinfo{person}{Thomas Unterthiner}, \bibinfo{person}{Andreas Mayr}, {and}
  \bibinfo{person}{Sepp Hochreiter}.} \bibinfo{year}{2017}\natexlab{}.
\newblock \showarticletitle{Self-normalizing neural networks}. In
  \bibinfo{booktitle}{\emph{Advances in neural information processing
  systems}}. \bibinfo{pages}{971--980}.
\newblock


\bibitem[\protect\citeauthoryear{Lee, Heo, Lee, Kim, and Jeong}{Lee
  et~al\mbox{.}}{2017}]%
        {lee2017applying}
\bibfield{author}{\bibinfo{person}{Sun-Ro Lee}, \bibinfo{person}{Min-Jae Heo},
  \bibinfo{person}{Chan-Gun Lee}, \bibinfo{person}{Milhan Kim}, {and}
  \bibinfo{person}{Gaeul Jeong}.} \bibinfo{year}{2017}\natexlab{}.
\newblock \showarticletitle{Applying deep learning based automatic bug triager
  to industrial projects}. In \bibinfo{booktitle}{\emph{Proceedings of the 2017
  11th Joint Meeting on Foundations of Software Engineering}}.
  \bibinfo{pages}{926--931}.
\newblock


\bibitem[\protect\citeauthoryear{Lou, Lin, Ding, Fu, Zhang, and Xie}{Lou
  et~al\mbox{.}}{2013}]%
        {lou2013software}
\bibfield{author}{\bibinfo{person}{Jian-Guang Lou}, \bibinfo{person}{Qingwei
  Lin}, \bibinfo{person}{Rui Ding}, \bibinfo{person}{Qiang Fu},
  \bibinfo{person}{Dongmei Zhang}, {and} \bibinfo{person}{Tao Xie}.}
  \bibinfo{year}{2013}\natexlab{}.
\newblock \showarticletitle{Software analytics for incident management of
  online services: An experience report}. In \bibinfo{booktitle}{\emph{2013
  28th IEEE/ACM International Conference on Automated Software Engineering
  (ASE)}}. IEEE, \bibinfo{pages}{475--485}.
\newblock


\bibitem[\protect\citeauthoryear{Lou, Lin, Ding, Fu, Zhang, and Xie}{Lou
  et~al\mbox{.}}{2017}]%
        {lou2017experience}
\bibfield{author}{\bibinfo{person}{Jian-Guang Lou}, \bibinfo{person}{Qingwei
  Lin}, \bibinfo{person}{Rui Ding}, \bibinfo{person}{Qiang Fu},
  \bibinfo{person}{Dongmei Zhang}, {and} \bibinfo{person}{Tao Xie}.}
  \bibinfo{year}{2017}\natexlab{}.
\newblock \showarticletitle{Experience report on applying software analytics in
  incident management of online service}.
\newblock \bibinfo{journal}{\emph{Automated Software Engineering}}
  \bibinfo{volume}{24}, \bibinfo{number}{4} (\bibinfo{year}{2017}),
  \bibinfo{pages}{905--941}.
\newblock


\bibitem[\protect\citeauthoryear{Matter, Kuhn, and Nierstrasz}{Matter
  et~al\mbox{.}}{2009}]%
        {matter2009assigning}
\bibfield{author}{\bibinfo{person}{Dominique Matter}, \bibinfo{person}{Adrian
  Kuhn}, {and} \bibinfo{person}{Oscar Nierstrasz}.}
  \bibinfo{year}{2009}\natexlab{}.
\newblock \showarticletitle{Assigning bug reports using a vocabulary-based
  expertise model of developers}. In \bibinfo{booktitle}{\emph{2009 6th IEEE
  international working conference on mining software repositories}}. IEEE,
  \bibinfo{pages}{131--140}.
\newblock


\bibitem[\protect\citeauthoryear{Murphy and Cubranic}{Murphy and
  Cubranic}{2004}]%
        {murphy2004automatic}
\bibfield{author}{\bibinfo{person}{G Murphy} {and} \bibinfo{person}{D
  Cubranic}.} \bibinfo{year}{2004}\natexlab{}.
\newblock \showarticletitle{Automatic bug triage using text categorization}. In
  \bibinfo{booktitle}{\emph{Proceedings of the Sixteenth International
  Conference on Software Engineering \& Knowledge Engineering}}. Citeseer.
\newblock


\bibitem[\protect\citeauthoryear{Potharaju, Chan, Hu, Nita-Rotaru, Wang, Zhang,
  and Jain}{Potharaju et~al\mbox{.}}{2015}]%
        {potharajuconfseer}
\bibfield{author}{\bibinfo{person}{Rahul Potharaju}, \bibinfo{person}{Joseph
  Chan}, \bibinfo{person}{Luhui Hu}, \bibinfo{person}{Cristina Nita-Rotaru},
  \bibinfo{person}{Mingshi Wang}, \bibinfo{person}{Liyuan Zhang}, {and}
  \bibinfo{person}{Navendu Jain}.} \bibinfo{year}{2015}\natexlab{}.
\newblock \showarticletitle{ConfSeer: Leveraging Support Knowledge Bases for
  Automated Misconfiguration Detection}.
\newblock \bibinfo{journal}{\emph{Proceedings of the VLDB Endowment}}
  \bibinfo{volume}{8}, \bibinfo{number}{12} (\bibinfo{year}{2015}),
  \bibinfo{pages}{1828--1839}.
\newblock


\bibitem[\protect\citeauthoryear{Potharaju, Jain, and Nita-Rotaru}{Potharaju
  et~al\mbox{.}}{2013}]%
        {potharaju2013juggling}
\bibfield{author}{\bibinfo{person}{Rahul Potharaju}, \bibinfo{person}{Navendu
  Jain}, {and} \bibinfo{person}{Cristina Nita-Rotaru}.}
  \bibinfo{year}{2013}\natexlab{}.
\newblock \showarticletitle{Juggling the jigsaw: Towards automated problem
  inference from network trouble tickets}. In
  \bibinfo{booktitle}{\emph{Presented as part of the 10th $\{$USENIX$\}$
  Symposium on Networked Systems Design and Implementation ($\{$NSDI$\}$ 13)}}.
  \bibinfo{pages}{127--141}.
\newblock


\bibitem[\protect\citeauthoryear{Rudolph, Ruiz, Mandt, and Blei}{Rudolph
  et~al\mbox{.}}{2016}]%
        {rudolph2016exponential}
\bibfield{author}{\bibinfo{person}{Maja Rudolph}, \bibinfo{person}{Francisco
  Ruiz}, \bibinfo{person}{Stephan Mandt}, {and} \bibinfo{person}{David Blei}.}
  \bibinfo{year}{2016}\natexlab{}.
\newblock \showarticletitle{Exponential family embeddings}. In
  \bibinfo{booktitle}{\emph{Advances in Neural Information Processing
  Systems}}. \bibinfo{pages}{478--486}.
\newblock


\bibitem[\protect\citeauthoryear{Vondrick, Khosla, Malisiewicz, and
  Torralba}{Vondrick et~al\mbox{.}}{2013}]%
        {vondrick2013hoggles}
\bibfield{author}{\bibinfo{person}{Carl Vondrick}, \bibinfo{person}{Aditya
  Khosla}, \bibinfo{person}{Tomasz Malisiewicz}, {and} \bibinfo{person}{Antonio
  Torralba}.} \bibinfo{year}{2013}\natexlab{}.
\newblock \showarticletitle{Hoggles: Visualizing object detection features}. In
  \bibinfo{booktitle}{\emph{Proceedings of the IEEE International Conference on
  Computer Vision}}. \bibinfo{pages}{1--8}.
\newblock


\bibitem[\protect\citeauthoryear{Williams}{Williams}{2019}]%
        {bigpandas}
\bibfield{author}{\bibinfo{person}{Maui Williams}.}
  \bibinfo{year}{2019}\natexlab{}.
\newblock \bibinfo{booktitle}{\emph{Tips for Modern NOCs – Alleviating
  Incident Routing Bottlenecks}}.
\newblock
\urldef\tempurl%
\url{https://www.bigpanda.io/blog/tips-for-modern-nocs-alleviating-incident-routing-bottlenecks/}
\showURL{%
Retrieved February 13, 2020 from \tempurl}


\bibitem[\protect\citeauthoryear{Xi, Yao, Xiao, Xu, and Lv}{Xi
  et~al\mbox{.}}{2019}]%
        {xi2019bug}
\bibfield{author}{\bibinfo{person}{Sheng-Qu Xi}, \bibinfo{person}{Yuan Yao},
  \bibinfo{person}{Xu-Sheng Xiao}, \bibinfo{person}{Feng Xu}, {and}
  \bibinfo{person}{Jian Lv}.} \bibinfo{year}{2019}\natexlab{}.
\newblock \showarticletitle{Bug Triaging Based on Tossing Sequence Modeling}.
\newblock \bibinfo{journal}{\emph{Journal of Computer Science and Technology}}
  \bibinfo{volume}{34}, \bibinfo{number}{5} (\bibinfo{year}{2019}),
  \bibinfo{pages}{942--956}.
\newblock


\bibitem[\protect\citeauthoryear{Xia, Lo, Wang, and Zhou}{Xia
  et~al\mbox{.}}{2013}]%
        {xia2013accurate}
\bibfield{author}{\bibinfo{person}{Xin Xia}, \bibinfo{person}{David Lo},
  \bibinfo{person}{Xinyu Wang}, {and} \bibinfo{person}{Bo Zhou}.}
  \bibinfo{year}{2013}\natexlab{}.
\newblock \showarticletitle{Accurate developer recommendation for bug
  resolution}. In \bibinfo{booktitle}{\emph{2013 20th Working Conference on
  Reverse Engineering (WCRE)}}. IEEE, \bibinfo{pages}{72--81}.
\newblock


\bibitem[\protect\citeauthoryear{Xie, Zhang, Yang, and Wang}{Xie
  et~al\mbox{.}}{2012}]%
        {xie2012dretom}
\bibfield{author}{\bibinfo{person}{Xihao Xie}, \bibinfo{person}{Wen Zhang},
  \bibinfo{person}{Ye Yang}, {and} \bibinfo{person}{Qing Wang}.}
  \bibinfo{year}{2012}\natexlab{}.
\newblock \showarticletitle{Dretom: Developer recommendation based on topic
  models for bug resolution}. In \bibinfo{booktitle}{\emph{Proceedings of the
  8th international conference on predictive models in software engineering}}.
  \bibinfo{pages}{19--28}.
\newblock


\bibitem[\protect\citeauthoryear{Xu, Huang, Fox, Patterson, and Jordan}{Xu
  et~al\mbox{.}}{2009}]%
        {xu2009detecting}
\bibfield{author}{\bibinfo{person}{Wei Xu}, \bibinfo{person}{Ling Huang},
  \bibinfo{person}{Armando Fox}, \bibinfo{person}{David Patterson}, {and}
  \bibinfo{person}{Michael~I Jordan}.} \bibinfo{year}{2009}\natexlab{}.
\newblock \showarticletitle{Detecting large-scale system problems by mining
  console logs}. In \bibinfo{booktitle}{\emph{Proceedings of the ACM SIGOPS
  22nd symposium on Operating systems principles}}. \bibinfo{pages}{117--132}.
\newblock


\end{thebibliography}

\end{document}